\documentclass[iop,revtex4]{emulateapj}

\usepackage{hyperref}
\usepackage{epsfig}
\usepackage{natbib}
\usepackage{graphicx}                % To format bibliographies.
\usepackage{lscape}
\usepackage{verbatim}              % Allows quoting source with commands.
\usepackage{amsmath,amsthm,amsfonts,amsopn,amssymb} % Some nice math packages.
\usepackage{longtable}
\usepackage{pdflscape}
\usepackage{afterpage}
\usepackage{rotating}					% Allows you to rotate figures, text and tables 
\usepackage{ulem}
\usepackage{epstopdf}  
\usepackage{multirow}
\usepackage{color}
\usepackage[left]{lineno}
\received{}
\accepted{}

\slugcomment{to appear in ApJ}
\shorttitle{}
\shortauthors{Wang}

\begin{document}
\title{Influence of Stellar Multiplicity On Planet Formation. II.  Planets Are Less Common in Multiple-Star Systems with Separations Smaller than 1500 AU}
\author{
Ji Wang\altaffilmark{1},
Debra A. Fischer\altaffilmark{1},
Ji-Wei Xie\altaffilmark{2},
David R. Ciardi\altaffilmark{3},
%Tabetha S. Boyajian\altaffilmark{2},
%Justin R. Crepp\altaffilmark{5},
%Megan E. Schwamb\altaffilmark{6,7},
%Chris Lintott\altaffilmark{8,9},
%Kian J. Jek\altaffilmark{10},
%Arfon M. Smith\altaffilmark{9},
%Michael Parrish\altaffilmark{9},
%Kevin Schawinski \altaffilmark{11},
%Joseph Schmitt\altaffilmark{2},
%Matthew J. Giguere\altaffilmark{2},
%John M. Brewer\altaffilmark{2},
%Stuart Lynn\altaffilmark{9},
%Robert Simpson\altaffilmark{8},
%Abe J. Hoekstra\altaffilmark{10},
%Thomas Lee Jacobs\altaffilmark{10},
%Daryll LaCourse\altaffilmark{10},
%Hans Martin Schwengeler\altaffilmark{10},
%Mike Chopin\altaffilmark{10}
} 
\email{ji.wang@yale.edu}
%%
%%% Some possible altaffiltext entries
%%\altaffiltext{1}{This publication has been made possible by the participation of 
%%more than 200,000 volunteers in the Planet Hunters project. Their contributions are 
%%individually acknowledged at http://www.planethunters.org/authors}
\altaffiltext{1}{Department of Astronomy, Yale University, New Haven, CT 06511 USA}
%\altaffiltext{2}{Department of Astronomy and Astrophysics, University of Toronto, Toronto, ON M5S 3H4, Canada}
\altaffiltext{2}{Department of Astronomy \& Key Laboratory of Modern Astronomy and Astrophysics in Ministry of Education, Nanjing University,
210093, China}
%\altaffiltext{4}{NASA Ames Research Center, M/S 244-30, Moffett Field, CA 94035, USA}
%\altaffiltext{5}{Bay Area Environmental Research Institute, Inc., 560 Third Street West, Sonoma, CA 95476, USA}
\altaffiltext{3}{NASA Exoplanet Science Institute, Caltech, MS 100-22, 770 South Wilson Avenue, Pasadena, CA 91125, USA}
%%\altaffiltext{6}{Department of Physics, Yale University, P.O. Box 208121, New Haven, CT 06520, USA}
%%\altaffiltext{7}{Yale Center for Astronomy and Astrophysics, Yale University, P.O. Box 208121, New Haven, CT 06520, USA}
%%\altaffiltext{8}{Oxford Astrophysics, Denys Wilkinson Building, Keble Road, Oxford OX1 3RH}
%%\altaffiltext{9}{Adler Planetarium, 1300 S. Lake Shore Drive, Chicago, IL 60605, USA}
%%\altaffiltext{10}{Planet Hunter}
%%\altaffiltext{11}{Institute for Astronomy, Department of Physics, ETH Zurich, Wolfgang-Pauli-Strasse 16, CH-8093 Zurich, Switzerland}

\begin{abstract}

Almost half of the stellar systems in the solar neighborhood are made up of multiple stars. In multiple-star systems, planet formation is under the dynamical influence of stellar companions, and the planet occurrence rate is expected to be different from that for single stars. There have been numerous studies on the planet occurrence rate of single star systems. However, to fully understand planet formation, the planet occurrence rate in multiple-star systems needs to be addressed. In this work, we {{infer}} the planet occurrence rate in multiple-star systems by measuring the stellar multiplicity rate for planet host stars. For a sub-sample of 56 $Kepler$ planet host stars, we use adaptive optics (AO) imaging and the radial velocity (RV) technique to search for stellar companions. The combination of these two techniques results in high search completeness for stellar companions. We detect 59 visual stellar companions to 25 planet host stars with AO data. {{Three stellar companions are within 2$^{\prime\prime}$, and 27 within 6$^{\prime\prime}$. We also detect 2 possible stellar companions (KOI 5 and KOI 69) showing long-term RV acceleration.}} After correcting for a bias against planet detection in multiple-star systems due to flux contamination, we find that planet formation is suppressed in multiple-star systems with separations smaller than 1500 AU. Specifically, we find that compared to single star systems, planets in multiple-star systems occur $4.5\pm3.2$, $2.6\pm1.0$, and $1.7\pm0.5$ times less frequently when a stellar companion is present at a distance of 10, 100, and 1000 AU, respectively. This conclusion applies only to circumstellar planets; the planet occurrence rate for circumbinary planets requires further investigation. 

\end{abstract}

%\keywords{Planets and satellites: detection - surveys}

\section{Introduction}

The majority of the stars in the solar neighborhood belong to multiple-star systems~\citep{Duquennoy1991, Fischer1992, Raghavan2010,Duchene2013}. In {{multiple-star systems}}, many planets have been detected. Some planets are detected in circumbinary orbits~\citep[P-type,][]{Dvorak1982}, where the planet orbits both stars~\citep[e.g., ][]{Doyle2011, Welsh2012,Schwamb2013}. Some others are detected in circumstellar orbits (S-type), where the planet orbits only one of the stars~\citep[e.g., ][]{Cochran1997,Eggenberger2004}. Compared to our statistical knowledge of planets around single stars~\citep{Cumming2008,Howard2010,Mayor2011,Wright2012, Mann2012,Dressing2013,Gaidos2013,Swift2013,Kopparapu2013,Petigura2013,Petigura2013b,Bonfils2013,Parker2013}, our understanding of planet formation in multiple-star systems is rather limited; the planet occurrence rate in multiple-star systems is still largely unconstrained.

%There are two avenues of studying planets in multiple-star systems: (1), to directly search for planets, and measure the planet occurrence rate in multiple star systems~\citep[e.g., ][]{Konacki2005, Eggenberger2007, Konacki2009}; (2) to search for stellar companions to planet host stars, and measure the stellar multiplicity rate for planet host stars~\citep[e.g., ][]{Luhman2002,Eggenberger2011}. The first avenue is straightforward, but it is prone to flux contamination of stellar companions, which affects measurement precision~\citep{Wright2012}. In comparison, the technical challenges for the second avenue are smaller; it is easier to search for a star than it is to search for a planet. The stellar multiplicity rate for planet host stars is an indirect measure of the planet occurrence rate in multiple-star systems. For example, a low stellar multiplicity rate for planet host stars, i.e., most planet host stars are single stars, may indicate that planet occurrence rate is low in multiple-star systems. 

Planets in multiple-star systems can be studied by either searching for planets in known multiple-star systems, or searching for stellar companions in known planetary systems. There have been a few studies to search for planets in known multiple-star systems~\citep[e.g., ][]{Konacki2005, Eggenberger2007, Konacki2009,Toyota2009}. However, this direct method is prone to flux contamination of stellar companions, which affects measurement precision~\citep{Wright2012}. In comparison, the technical challenges are dramatically reduced for detecting stellar companions in known planetary systems; it is easier to search for a star than it is to search for a planet. Determination of the stellar multiplicity rate for planet host stars solves the inverse problem of measuring the planet occurrence rate in multiple-star systems~\citep{Wang2014}. If planet host stars are rarely in multiple-star systems, this would indicate a low planet occurrence rate in these systems.

There have been numerous studies which have measured the stellar multiplicity rate of planet host stars. Most of these studies used imaging techniques, such as the adaptive optics (AO) imaging~\citep{Luhman2002,Patience2002,Eggenberger2007,Eggenberger2011, Adams2012, Adams2013,Law2013,Dressing2014}, Lucky Imaging~\citep{Daemgen2009, Ginski2012, LilloBox2012,Bergfors2013,LilloBox2014}, speckle imaging~\citep{Horch2012,Kane2014}, wide field imaging~\citep{Mugrauer2007,Mugrauer2009}, HST imaging~\citep{Gilliland2014}, and other techniques~\citep{Raghavan2006,Raghavan2010,Roell2012}. These studies have mostly reached similar conclusions that the stellar multiplicity rate of planet host stars is lower than or comparable to that for field stars in the solar neighborhood. Among these studies, some focused on stars hosting planets detected in ground-based RV or transiting surveys~\citep{Luhman2002,Patience2002,Eggenberger2004,Raghavan2006,Mugrauer2007,Eggenberger2007,Daemgen2009,Mugrauer2009,Raghavan2010,Roell2012,Ginski2012,Bergfors2013,Knutson2013}. However, the bias of ground-based planet surveys is difficult to assess due to an unknown threshold for excluding multiple-star systems. 

In comparison, the $Kepler$ mission~\citep{Borucki2011,Batalha2013,Burke2014} did not strongly bias against multiple-star systems: (1), the low-angular-resolution $Kepler$ Input Catalog images~\citep{Brown2011} did not {{reveal close binaries}}; (2), multiple-star systems (e.g., eclipsing binaries) received continued observation after detection. Therefore, the bias of ground-based surveys is not a major concern for studies of $Kepler$ planet host stars~\citep{LilloBox2012,Adams2012,Horch2012,Adams2013,Law2013,Dressing2014,Kane2014,Gilliland2014,LilloBox2014}. However, there is a detection bias against transiting planets in multiple-star systems. The transit depth is shallower due to the additional flux from stellar companions, which makes planet detection more difficult. This bias has to be taken into consideration when measuring the stellar multiplicity rate for planet host stars. 

It is commonly accepted that planet formation may be disrupted in multiple-star systems with small separations (e.g., $\sim$10-200 AU). This is supported by both simulations~\citep{Thebault2006,JangCondell2007,Quintana2007,Paardekooper2008,Kley2008,Xie2010,Thebault2011} and observations~\citep{Desidera2007, Bonavita2007, Kraus2012}. Therefore, surveys for gravitationally-bound stellar companions around planet host stars provides the best path for understanding planet formation in multiple-star systems. High-resolution imaging techniques are efficient for separations greater than 0$^{\prime\prime}$.1, and spectroscopic measurements are efficient at detecting stellar companions at smaller separations.

To carry out this work, we select a sample of 56 stars hosting planet candidates from the $Kepler$ mission to search for potential stellar companions using the RV and the AO imaging techniques. The RV technique is sensitive to short-period stellar companions, and the AO technique is sensitive to those further out. The combination of these two techniques is sensitive to a larger range of semi-major axes, and results in a survey with much higher completeness than previous studies. We consider the detection bias against transiting planets in multiple-star systems, and correct for this when calculating the stellar multiplicity rate for planet host stars. We emphasize that we only consider planets in S-type circumstellar orbits.

The paper is organized as follows. We describe our sample and the sources for their RV and AO data in \S \ref{sec:Sample}. In \S \ref{sec:Search}, we present the analyses of available data: searching for stellar companions to planet host stars using the RV and AO techniques. In \S \ref{sec:planet_frequency}, we introduce a method of correcting for detection bias against planets in multiple-star systems, and apply it to the measurement of stellar multiplicity rate for planet host stars. We then calculate the planet occurrence rate for single and multiple star systems by comparing their multiplicity rates. Discussion and summary are given in \S \ref{sec:Summary}.

\section{Sample Description and Data Sources}
\label{sec:Sample}

RV and AO data are provided by the $Kepler$ Community Follow-up Observation Program\footnote{https://cfop.ipac.caltech.edu} (CFOP). Since RV data are critical for probing stellar companions on close orbits, we select 56 $Kepler$ Objects of Interest (KOIs) with at least 3 RV data points, for which the long-term RV acceleration due to a stellar companion may be measured. The RV data were taken with the HIRES instrument~\citep{Vogt1994} at the Keck I telesope, and reported in~\citet{Marcy2014}.

The AO data for these 56 KOIs were taken at different observatories including Keck, MMT, Gemini, Lick, Palomar, and {{WIYN}}. A summary of the sample and data sources is given in Table \ref{tab:stellar_params}. Information on KOIs is provided by the NASA Exoplanet Archive~\citep[NEA,][]{Huber2014}\footnote{http://exoplanetarchive.ipac.caltech.edu/}. All the stars in our sample are solar-type stars with effective temperature ($T_{\rm{eff}}$) in the range between 4725 K and 6300 K, and surface gravity (log $g$) in the range of 3.9 and 4.7. There are 27 (48\% of the sample) multi-planet systems. 

\section{Detections and Constraints on Stellar Companions to Planet Host Stars}
\label{sec:Search}

{{We use three techniques to detect and constrain stellar companions around planet host stars: the RV technique (\S \ref{sec:rv}), the AO imaging technique (\S \ref{sec:ao}), and the dynamical analysis (DA, \S \ref{sec:dynamical}). These three techniques are complementary and sensitive to different parts of parameter space. The RV technique is sensitive to close-in companions with small to intermediate mutual inclinations with respect to the planet orbital planes; the DA technique is sensitive to companions at large mutual inclinations; and the AO imaging technique is sensitive to stellar companions at larger separations. We will discuss in this section how we use these techniques to detect stellar companions and constrain their presence.}}

\subsection{RV Detections and Completeness}
\label{sec:rv}

We use the Keplerian Fitting Made Easy (KFME) package~\citep{Giguere2012} to analyze the RV data. The procedure are described as follows. We first calculate the root mean square (RMS$_1$ in Table \ref{tab:rv}) of the the RV data. If RMS$_1$ is five times higher than the median reported RV uncertainties, $\delta v$, then we mark a variability flag. For systems marked with variability flags, we first fit the RV data with a linear trend, or a long-period orbit due to a non-transiting object. The systems with a significant linear trend (3-$\sigma$) or the signal of an additional non-transiting object will be marked with a slope flag or a non-transiting component flag. The linear trend or the long-period orbit is then removed. 

For the RV residuals after removing the linear trend or the long-period orbit, and the RVs for systems with no variability flag, we consider two cases. First, if the system has only one detected KOI, then we fit the RV data with a circular Keplerian orbit allowing only the RV semi-amplitude to change. If the resulting RMS (RMS$_2$) is smaller than RMS$_1$, then RMS$_2$ is used in subsequent analyses; otherwise RMS$_1$ is used. {{Second, if the system has multiple KOIs, then we choose the one KOI causing the largest RV variation. When deciding which planet in the KOI causes the largest RV variation, we assume the planet mass-radius relationship from~\citet{Lissauer2011}, and calculate the nominal RV amplitude for each planet. Then we fit the RV data with a circular Keplerian orbit allowing only the RV semi-amplitude to change.}} The minimum of RMS$_1$ and RMS$_2$ is used in subsequent analyses. In the above process, fitting eccentric orbit does not significantly change the RMS. In addition, including more KOIs for multi-planet systems does not always help to reduce the RMS because of large RV measurement uncertainty relative to the small RV signals of small planets. 

Among 56 stars, only one shows a long-term RV slope: KOI 69 has a RV linear trend of $12.2\pm0.2$ m$\rm{s}^{-1}\ \rm{yr}^{-1}$. During 4.1 years of RV measurements, there is no sign of deviation from the linear trend, indicating that the companion is at least 4 AU away. More discussion regarding this system will be given in \S \ref{sec:ao} after considering the AO data. Six systems have non-transiting objects revealed by the RV data: KOI 5, KOI 104, KOI 148, KOI 244, KOI 246, and KOI 1442. The latter five are known non-transiting planets~\citep{Marcy2014}. KOI 5 shows a parabolic acceleration, but the period of the non-transiting object is unconstrained. We will discuss KOI 5 more in \S \ref{sec:ao} with the addition of AO data. {{For {{}15} cases, RMS is still 5 times higher than the reported RV measurement uncertainty after considering a non-transiting object, or the KOI planet dominating the RV variability. The ``excessive" RV variability may be attributed to the following factors or their combinations: very limited number of RV data points, an underestimated RV measurement uncertainty, excessive stellar activity, and additional stellar or planetary components. We find that 12 out of the 15 KOIs with ``excessive" RV variation have fewer than 21 RV measurements, which is the median number of RV measurements for the 56 KOIs in our sample. Seven of them have fewer than 10 RV measurements. The limited number of RV measurements would result in an improper RV orbital fitting, which leads to a higher RMS. In addition, RV jitter is not considered in the reported RV uncertainty in Table \ref{tab:rv}. After considering a typical RV jitter of 1-3 m$\cdot$s$^{-1}$ for Kepler stars with RV measurements~\citep{Marcy2014}, 10 out of the 15 KOIs with ``excessive" RV variation have less than 5 times of the RV uncertainty. KOI 22 remains the only KOI in our sample with ``excessive" RV variation that cannot be explained by either the limited number of RV measurements or stellar activity.}} 

\begin{figure}[htp]
\begin{center}
\includegraphics[angle=0, width=0.45\textwidth]{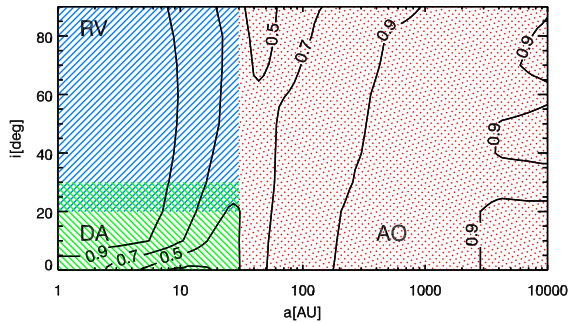} 
\caption{Median completeness contours for 3 techniques used to detect and constrain stellar companions to planet host stars. These 3 techniques are sensitive to different part of the $a-i$ parameter space. The radial velocity (RV) technique is sensitive to stellar companions within $\sim$30 AU and with small or intermediate mutual inclinations to planet orbital planes (blue hatched region). The dynamical analysis (DA) technique is sensitive to a similar range of separations but larger mutual inclinations (green hatched region). The adaptive optic (AO) technique is sensitive to stellar companions at wider orbits (red dotted region). The combination of these 3 techniques contributes to a survey of stellar companions with high completeness. 
\label{fig:RV_AO_DA_completeness}}
\end{center}
\end{figure}

We study the completeness of searching for stellar companions by simulations following the subsequent procedures. We first define a parameter space, $a-i$ space, where $a$ is the separation of a companion star, and $i$ is the mutual inclination of the sky plane and the companion star orbital plane. We divide the parameter space into many fine grids ($\Delta a=0.5$ AU, $\Delta i=10^\circ$). For each star, we simulate 1000 companion stars on each grid, and count how many simulated companion stars are detected given the time baseline, observation epochs, and measurement uncertainties of the RV data. Specifically, we generate a synthetic RV data set for each of the simulated companion stars. Observation epochs and measurement uncertainties remain the same as the original RV data. If the RMS of the synthetic RV data is 3 times larger than the observed RV RMS, i.e., the smaller of RMS$_1$ and RMS$_2$, then we count the simulated stellar companion as a detection. The separation and mass ratio distributions of simulated stellar companions follow the normal distributions reported in~\citet{Duquennoy1991}, i.e., $\log_{10}a=1.49,\ \sigma_{\log_{10}a}=1.54;\ q=m2/m1=0.23,\ \sigma_q=0.42$. We use the median orbital eccentricity for binary stars~\citep[$e=0.4$,][]{Duquennoy1991} and a random periastron distribution in simulations. The median completeness contours are shown in Fig. \ref{fig:RV_AO_DA_completeness}. RV completeness drops to below 50\% as separations become larger than 30 AU. 

In summary, RV observations of 56 stars reveal 7 non-transiting companions, 5 of these are previously-reported planets~\citep{Marcy2014}. Orbits of the other two are unconstrained because of limited RV baselines. KOI 5 shows a parabolic RV acceleration, and KOI 69 shows a linear RV trend of $12.2\pm0.2$ m$\rm{s}^{-1}\ \rm{yr}^{-1}$. The nature of these two companions will be discussed more in the following section. 

\subsection{AO Detections and Completeness}
\label{sec:ao}

{{The RV variation of most of stellar companions at larger separations is difficult to measure because of the long periodicity. However, the AO imaging technique is more effective in constraining stellar companions at larger separations. We will discuss in the following part how we detect and characterize stellar companions based on AO images.}}

\subsubsection{Contrast Curve}
\begin{figure}[htp]
\begin{center}
\includegraphics[angle=0, width= 0.45\textwidth]{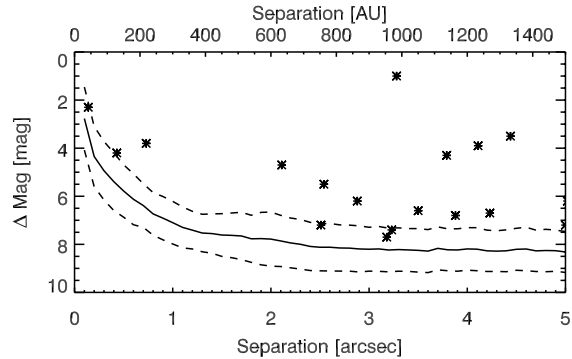} 
\caption{Median contrast curve for the AO images. Dashed lines are 1-$\sigma$ deviation of the contrast curve. Detections within 5$^{\prime\prime}$ are shown as asterisks. Physical projected separation in AU is calculated assuming the average distance of the sample, i.e., 300 pc. When analyzing the detection completeness, each star in our sample is treated individually for the observation band in which the AO image was taken. A total of 59 visual companions around 25 planet host stars are detected (Table \ref{tab:AO_params}). 
\label{fig:AO_contrast}}
\end{center}
\end{figure}

The contrast curve of an image provides detection thresholds for detecting faint companions around a star. The procedures of calculating the contrast curve are described as follows. We define a series of concentric annuli, centered on the star, for which we calculate the median and the standard deviation of flux for pixels within these annuli. We use the value of 5 times the standard deviation above the median as the 5-$\sigma$ detection limit. The contrast curve is the 5-$\sigma$ detection limit as a function of the radii of concentric annuli. The median contrast curve and the 1-$\sigma$ deviation of the AO images we use in this paper are shown in Fig. \ref{fig:AO_contrast}, where each pixel is converted into angular separation based on plate scale of each instrument: 0$^{\prime\prime}$.010/pixel for Keck NIRC2~\citep{Wizinowich2000}, 0$^{\prime\prime}$.011/pixel for Gemini DSSI\footnote{http://www.gemini.edu/sciops/instruments/dssi-speckle-camera-north}, 0$^{\prime\prime}$.019/pixel or 0$^{\prime\prime}$.038/pixel for MMT ARIES~\citep{Sarlot1999}, 0$^{\prime\prime}$.025/pixel for Palomar PHARO~\citep{Hayward2001}, 0$^{\prime\prime}$.075/pixel for Lick IRCAL~\citep{Lloyd2000}, 0$^{\prime\prime}$.017-0$^{\prime\prime}$.018/pixel for WIYN DSSI~\citep{Horch2009}, and 0$^{\prime\prime}$.043/pixel for Palomar Robo-AO~\citep{Law2013}.

\subsubsection{Distance Estimation}
In order to obtain the physical projected separation between detected companions and the central stars, we need to estimate the distance. The distance of a star can be measured with the distance modulus and an estimation of extinction. The extinction estimation in the $V$ band ($A_V$) is obtained from the Mikulski Archive for Space Telescopes (MAST)\footnote{https://archive.stsci.edu/}. {{Details of $A_V$ estimation can be found in \S 6 and \S 7 in~\citet{Brown2011}}}. The distance modulus is the magnitude difference between the apparent magnitude and the absolute magnitude in the $V$ band. The apparent $V$ magnitude is calculated based a conversion from $g^\prime$ and $r^\prime$ magnitudes~\citep{Smith2002}. The absolute $V$ magnitude is estimated with the Yale-Yonsei (Y2) stellar evolution model~\citep{Demarque2004}: with $T_{\rm{eff}}$, log $g$, age, and [Fe/H] measured from spectroscopic and/or asteroseismic observations, the absolute $V$ magnitude can be estimated from the Y2 interpolator. For stars with an unknown $A_V$, which is the case for 7 stars, we use the distance modulus in $K$ band to estimate the distance with the assumption that $K$ band extinction is much smaller than $V$ band for $Kepler$ stars. {{Distances for KOIs with visual stellar companion detections are provided in Table \ref{tab:AO_params}}}. 

\subsubsection{Detection and Completeness}

Based on the images from the CFOP, we detect a total of 59 visual stellar companions around 25 planet host stars (Table \ref{tab:AO_params}). Fourteen stars (25\%) have stellar companions within a 5$^{\prime\prime}$ radius. The closest companion has a projected separation of 40.9 AU (0$^{\prime\prime}$.14) from KOI 5. 

The 56 stars in our sample have an average distance of $\sim$300 pc. Given the contrast curve shown in Fig. \ref{fig:AO_contrast}, the search for stellar companions closer than $\sim$40 AU and low-mass stars ($\Delta$ Mag $>$ 8) is not complete. We therefore conduct simulations to evaluate the completeness of the AO survey. Similar to the RV completeness simulations in \S \ref{sec:rv}, we artificially generate 1000 companion stars at each predefined grid in the $a-i$ parameter space. If the contrast ratio ($\Delta$ Mag) between a simulation star and the central star is smaller than the value given by AO 5-$\sigma$ contrast curve, then we record it as a detection. {{Note that the contrast curves used in simulations are those calculated for each individual star in the observed band rather than the median contrast curve shown in Fig. \ref{fig:AO_contrast}.}} The AO completeness contours (median of 56 stars) are plotted in Fig. \ref{fig:RV_AO_DA_completeness}. From this plot, we show that the AO completeness is less than 50\% for separations smaller than $\sim$40 AU. At smaller separations, the RV technique becomes a much more efficient way of detecting stellar companions. 

\subsubsection{KOI 5}
\label{sec:koi05}

\begin{figure}[htp]
\begin{center}
\includegraphics[angle=0, width= 0.45\textwidth]{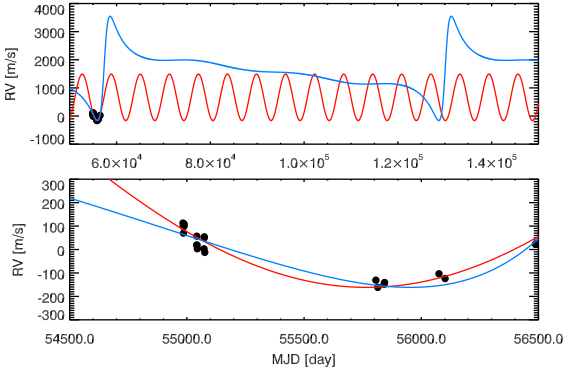} 
\caption{Two possible scenarios for the observed RV acceleration of KOI 5. Black dots are current RV data. Blue line shows a case in which the RV acceleration is caused by the AO detected companion with a 40.9 AU projected separation (i.e., 0$^{\prime\prime}.14$ angular separation). A highly eccentric orbit ($e=0.92$) is required to reasonably fit the RV data. Red line shows another case in which a 0.08 M$_\odot$ companion on a circular orbit with a 7 AU separation causes the RV acceleration. More RV data with a longer baseline are required to determine the nature of the companion causing the RV acceleration of KOI 5. Top panel shows a large time range, and the bottom panel shows a zoom-in plot to a time range with RV data. 
\label{fig:K00005}}
\end{center}
\end{figure}

KOI 5 has a parabolic RV acceleration indicating a distant companion, but the orbit of this companion is unconstrained given only $\sim$4 years' observation and the poor phase coverage. There are many possible orbital solutions given the current RV data. Fig. \ref{fig:K00005} shows two examples. If the RV acceleration is caused by the stellar companion detected by the AO imaging, then it requires a highly eccentric orbit ($e=0.92$) to reasonably fit the RV data. We estimate the mass of the AO detected stellar companion to be $\sim$0.5 $M_\odot$ based on its differential magnitude in the $K$ band~\citep{Kraus2007}. Alternatively, the observed RV acceleration can be explained by a stellar companion (0.08 $M_\odot$) at 7 AU separation on a circular orbit. Any solutions with separations smaller than 7 AU should involve companions that fall into sub-stellar mass regime. Therefore, we conclude that a stellar companion may exist around KOI 5, but with a separation larger than 7 AU (i.e., $0^{\prime\prime}.024$ angular separation). 

\subsubsection{KOI 69}
\label{sec:koi69}

\begin{figure}[htp]
\begin{center}
\includegraphics[angle=0, width= 0.45\textwidth]{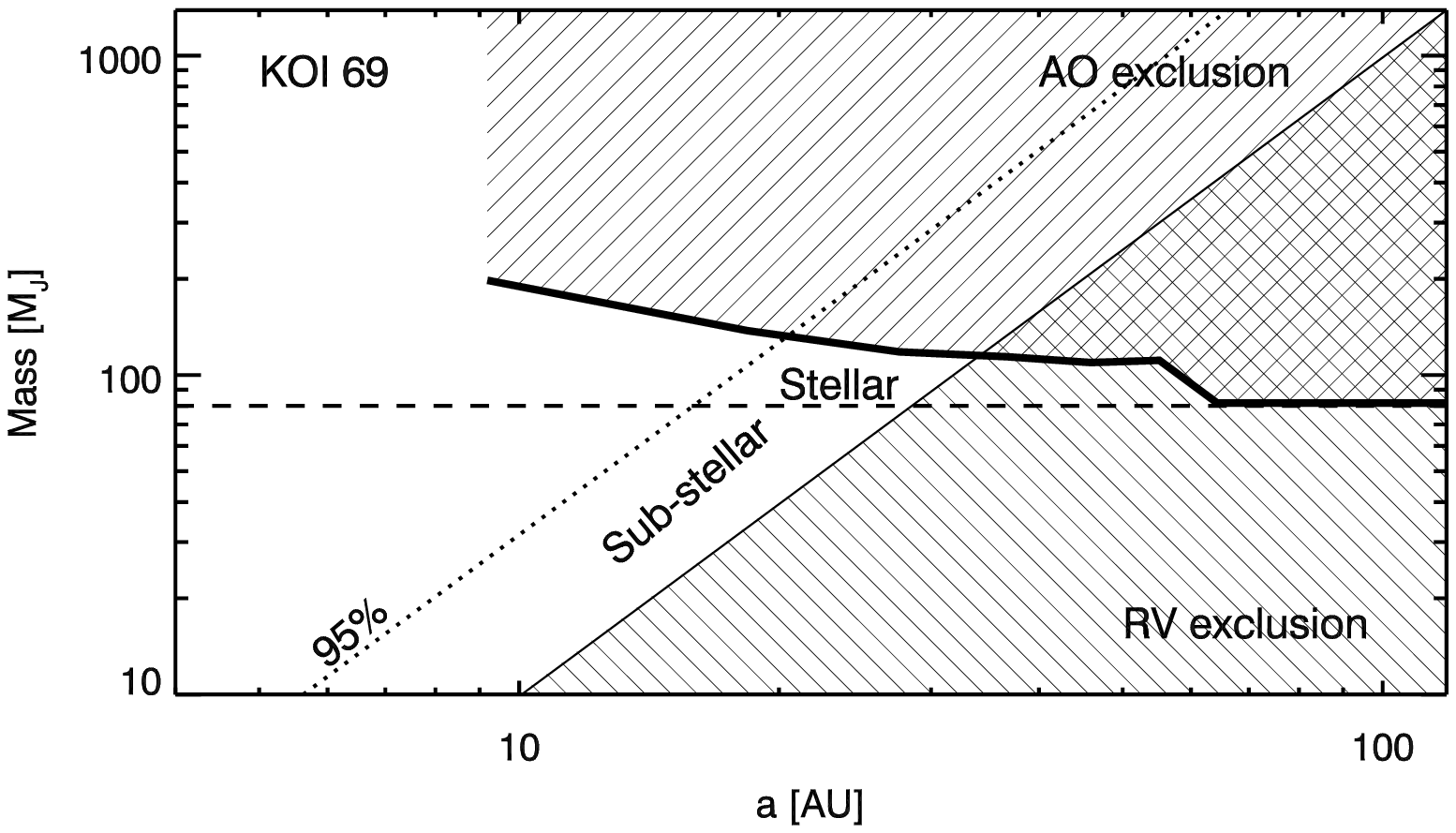} 
\caption{Parameter space for the companion to KOI 69 revealed by a RV linear trend. The region below the solid straight line is excluded because companions falling into this parameter space are not massive and close enough to produce the observed RV linear trend. The solid thick line represents the contrast curve. Any stellar companions above the line would have been detected, so the parameter space above the solid thick line is also excluded. The dashed line divides the stellar regime and sub-stellar regime. The dotted line represents the 95 percentile of the solutions given the linear RV trend, i.e., 95\% of the solutions should fall in between the solid straight line and the dotted line assuming a random orbital orientation of the companion. If the linear RV trend is caused by a stellar companion, then it is most likely the case that the separation is between 15.5 and 33.0 AU, as confined in a parameter space noted with ``Stellar".
\label{fig:K00069}}
\end{center}
\end{figure}

KOI 69 shows a linear RV trend of $12.2\pm0.2$ m$\rm{s}^{-1}\ \rm{yr}^{-1}$, which can be caused either by a more distant star or a closer sub-stellar object. Fig. \ref{fig:K00069} shows possible parameter space for this companion. RV data exclude any companions below the straight solid line because they are not massive enough to cause the trend. Although AO data shows non-detection for KOI 69, the AO contrast curve can put constraint on any bright stellar objects which would have been detected. {{After considering the constraints from AO and RV observations, if the companion causing the RV linear trend is a star, it is mostly likely to lie between 15.5 and 33.0 AU (i.e., $0^{\prime\prime}.18$ and $0^{\prime\prime}.38$ in angular separation), and its mass cannot exceed 102 Jupiter mass (2-$\sigma$). If the companion mass is in the sub-stellar regime, its mass and separation is confined to a parallelogram marked as ``Sub-stellar" in Fig. \ref{fig:K00069}. The 4 vertices of the parallelogram are (5.5 AU, 10.0 $M_J$), (9.8 AU, 10.0 $M_J$), (27.6 AU, 80.0 $M_J$), and (15.5 AU, 80.0 $M_J$).}}

\subsubsection{Visual Companions Association}
We detect 59 visual companions around 25 planet host stars, but we do not know whether they are associated or bonded to the host stars.~\citet{LilloBox2012} estimated $\sim$35\%-53\% of visual companions are bonded to the primary stars within 3$^{\prime\prime}$, and this ratio decreases with increasing angular separations. Therefore, a non-negligible fraction of visual companions we detect are in fact unassociated to primary stars, which will decrease the stellar multiplicity rate for planet host stars. 

For {{}12} visual companions with multi-band detections (i.e., $J$ and $K$ band), we test if they are physically associated with their primary stars. The procedures of the test are described as follows. We calculate the $J$-$K$ colors of visual companions based on the differential magnitudes in Table \ref{tab:AO_params} and the $J$-$K$ colors of primary stars from the NEA. From their $J$-$K$ colors, we interpolate for the absolute $K$ band magnitude of companion stars based on Table 5 in~\citet{Kraus2007}. With the absolute $K$ band magnitudes and the apparent $K$ band magnitudes, we calculate the distances of the visual companion stars, and check whether they are consistent with the distances of the primary stars. {{If the color-determined distance for the companion is 1 $\sigma$ different from the distance of the KOI as reported in Table \ref{tab:AO_params}, we reject the physical association between the KOI and the visual companion star. We find inconsistent distance for 6 out of 12 visual companions. The 6 companions include one for KOI 87 (at $5^{\prime\prime}.49$ separation, $d=2.9\pm2.2\ \rm{kpc}$), one for KOI 153 (at $6^{\prime\prime}.17$ separation, $d=52\pm44\ \rm{kpc}$), one for KOI 244 (at $8^{\prime\prime}.40$ separation, $d=33\pm20\ \rm{pc}$), and all three for KOI 377 ($d=3.8\pm2.2\ \rm{kpc}$, $d=204\pm163 \rm{kpc}$, and $d=412\pm330\ \rm{kpc}$).}}   

\subsection{Dynamical Analysis}
\label{sec:dynamical}
In addition to constraints from RV and AO data, more constraints of potential stellar companions can be put on multi-planet systems~\citep{Wang2014}. There are 27 (48\% of the sample) multi-planet systems in our sample for which we can apply the dynamical analysis (DA). The DA technique makes use of the co-planarity of multi-planet systems discovered by the $Kepler$ mission~\citep{Lissauer2011}. A stellar companion with high mutual inclination to the planetary orbits would have perturbed the orbits and significantly reduce the co-planarity of planetary orbits, and hence the probability of multi-planet transits. Therefore, the fact that we see multiple planet transiting helps to exclude the possibility of a highly-inclined stellar companion. The DA is complementary to the RV technique because it is sensitive to stellar companions with large mutual inclinations to the planetary orbits. The parameter space the DA is sensitive to is shown in Fig. \ref{fig:RV_AO_DA_completeness}.  

\subsection{Combining Results From Different Techniques}
\label{sec:comb}
%Three techniques of detecting and constraining stellar companions around planet host stars have been discussed in previous sections. The RV, AO, and DA techniques are complementary and sensitive to different parts of the $a-i$ parameter space (Fig. \ref{fig:RV_AO_DA_completeness}). The RV technique is sensitive to close-in companions with small to intermedium mutual inclinations with respect to the planet orbital planes; the DA technique is sensitive to companions at large mutual inclinations; and the AO imaging technique is sensitive to stellar companions at larger separations. 

For the RV and AO observations, detection completeness contours are calculated based on simulations given the observational constraints, such as the time baseline, cadence, measurement uncertainties, and the contrast curve. For the DA technique, numerical integrations give the fraction of time when multiple planets can stay with small mutual inclinations ($<5^\circ$) so that multiple transiting planets can be observed~\citep{Wang2014}. Note that the DA technique works only for systems with multiple planets, which account for 48\% of the sample. For systems with a single transiting planet, no constraint can be given by the DA technique. We denote $c_{\rm{RV}}$, $c_{\rm{AO}}$ and $c_{\rm{DA}}$ as the completenesses at a given point in the $a-i$ parameter space, overall completeness $c$ is equal to $1-(1-c_{\rm{RV}})\times(1-c_{\rm{AO}})\times(1-c_{\rm{DA}})$. We note that the calculation assumes each technique is independent and uniquely sensitive to a certain portion of the parameter space. This is generally the case since the RV technique completeness drops quickly beyond $\sim$30 AU, where the AO technique sensitivity is high. {{Similarly, the RV and DA techniques and the DA and AO techniques have little overlap in sensitivity parameter space.}} The overall completeness may be overestimated at the transition space, such as $a=30$ AU (for RV and AO) and $i=20^\circ$ (for RV and DA), because stellar companions falling into this parameter space can be detected by multiple techniques and thus the techniques become correlated. We also try another way of combining results from different techniques, in which we use the maximum completeness as the overall completeness. This approach assumes multiple techniques are correlated, however, it does not  significantly change the conclusions in this paper.

The completeness is then integrated over the $a-i$ parameter space. We assume a log-normal distribution for $a$~\citep{Duquennoy1991, Raghavan2010}, random distribution of $i$ for systems with only one transiting planet, and the $i$ distribution from~\citet{Hale1994} for systems with multiple transiting planets. The treatment for multiple transiting planet systems is detailed in~\citet{Wang2014}, i.e., a coplanar distribution for stellar companions within 15 AU, a random $i$ distribution for stellar companions beyond 30 AU, and a mixture $i$ distribution at intermediate separations between 15 and 30 AU. 

\section{Planet Occurrence Rate and Stellar Multiplicity Rate}
\label{sec:planet_frequency}

\subsection{Detection Bias Against Planets in Multiple-Star Systems}
\label{sec:bias}
Planets in multiple-star systems are more difficult to find using the transiting method because of flux contamination. We discuss how this bias against planet detection in multiple-star systems can be quantified. For the $Kepler$ mission, it is a necessary condition to become a planet candidate that the signal to noise ratio (S/N) should be higher than 7.1~\citep{Jenkins2010}. S/N can be calculated using the following equation:
\begin{equation} 
\label{eq:snr}
S/N=\frac{\delta}{CDPP_{\rm{eff}}}\sqrt{N_{\rm{transits}}},
\end{equation}
where $\delta$ is the transit depth, $CDPP_{\rm{eff}}$ is the effective combined differential photometric precision~\citep{Jenkins2010b}, a measure of photometric noise, and $N_{\rm{transits}}$ is the number of observed transits. We use a planet in a binary system as an example to calculate the transit depth:
\begin{equation} 
\label{eq:delta}
\delta=\frac{R^2_{\rm{PL}}}{R^2_{\ast}}\frac{F_\ast}{F_\ast+F_c},
\end{equation}
where $R_{\rm{PL}}$ is planet radius, $R_\ast$ is the radius of the star that the planet is transiting, $F$ denotes flux, and subscript $\ast$ and $c$ indicate the planet host star and the contaminating star, respectively. Two cases are considered for the above equation. First, if the planet transits the primary star, the transit depth is diluted by a factor of 2 at most, when $F_\ast$ and $F_c$ are identical. Second, if the planet transits the secondary star, the transit depth dilution effect due to flux contamination can be much larger than 2 even after considering the {{increase in the transit depth}} from a reducing $R_\ast$ in the first term of the equation. For an example of a solar-type star and a late-type M dwarf pair, the gain of a reducing $R_\ast$ can be a factor of 100 at most, but the flux ratio between the two stars can easily exceed $10^4$ in the $Kepler$ band. 

Therefore, we conduct simulations to quantify the detection bias against planets in binary star systems. For each KOI, we choose the one planet that gives the highest S/N. We add a companion star in the system and calculate the S/N in the presence of flux contamination for two cases: planet transiting the primary star and planet transiting the secondary star. {{In both cases, we assume the same period and transit duration from the NEA so that $CDPP_{\rm{eff}}$ and $N_{\rm{transits}}$ in Equation \ref{eq:snr} are the same, and flux contamination (see Equation \ref{eq:delta}) is the only factor that determines whether a planet is detected in the presence of a companion star. }}If the S/N is higher than 7.1, then the planet can still be detected by $Kepler$, but with a lower significance. We randomly assign a stellar companion (secondary star) to a KOI (primary star) and repeat this procedure 1000 times for both the primary and the secondary star. We record {{the fraction of planet detections}}, $\alpha$, which will be used in correcting for the bias of detecting planets in multiple-star systems (Table \ref{tab:Bias_params}). The median value of $\alpha$ for 56 stars in our sample is 0.89, implying that the detection bias is not severe, but certainly not negligible.  

In the simulations, we use the stellar parameters from the NEA for the primary star. When generating a stellar companion in the simulations, we assume the mass ratio distribution follows the normal distribution given in~\citet{Duquennoy1991}. The radius of the secondary star is calculated using a stellar mass-radius relationship~\citep{Feiden2012}. Estimation of stellar flux for both primary and secondary stars are based on Table 5 in~\citet{Kraus2007}. We calculated $CDPP_{\rm{eff}}$ by interpolating between 3, 6, and 12-hour $CDPP$s based on transit duration. 

\subsection{Distinguishing Planets in Single and Multiple-Star Systems}
The $Kepler$ mission has provided us with a large sample of planet candidates. However, we do not know whether the planet host stars are in single or multiple stellar systems. Distinguishing planets in single and multiple-star systems allows us to separately calculate the planet occurrence rate for these two types of stars, and to understand planet formation in different stellar environments~\citep{Wang2014}. Follow-up observations are critical in identifying additional stellar companions in planetary systems. Even in the case of non-detection, with RV, AO, and DA techniques, we can calculate the probability of a star being in a multiple-star system based on the completeness study. For example, if the overall completeness for a companion detection is 80\% and the stellar multiplicity rate is 46\%~\citep{Raghavan2010}, then the probability of the star having an undetected companion (or being a multiple-star) is (100\%-80\%)$\times$46\%=0.092. Following this procedure, we calculate the number of multiple-stars $N_M$ and the number of single stars $N_S$. Since $N_M$ and $N_S$ are the sums of probabilities, they will not necessarily be integers:
\begin{equation} 
\label{eq:nmns}
N_M=\sum\limits_{i=1}^{n} [p_M(i)/\alpha(i)],\ N_S=\sum\limits_{i=1}^{n} [1-p_M(i)],
\end{equation}
where $n$ is the total number of stars in the sample, $p_M(i)$ is the probability of the $i_{\rm{th}}$ star being a multiple-star system, $\alpha(i)$ is the correction factor for the detection bias for planets in multiple-star systems. {{The above equation is similar to Equation 6 in~\citet{Wang2014} except for the correction factor $\alpha$}}. Note that there is an implicit correction factor for single stars in Equation \ref{eq:nmns}. However, the correction factor for single stars is 1. If a stellar companion is detected for a KOI, then $p_M$ is assigned to 1, and $\alpha$ is also assigned to 1 because no bias exists in this case since a planet has already been detected in a multiple-star system. {{For an AO detected stellar companion, setting $p_M$ to 1 is an overestimation because the physical association of visual stellar components is not yet established. Therefore, the stellar multiplicity rate that will be subsequently determined is an upper limit.}}

We then define $f$ as the fraction of stars with planets, $f$ can be separated into two components:
\begin{equation} 
\label{eq:fre_all}
f={(1-{\rm{MR}})\times f_S+{\rm{MR}}\times f_M},
\end{equation}
where $\rm{MR}$ is the {{global}} stellar multiplicity rate, $f_S$ and $f_M$ are the fraction of stars with planets for single and multiple-star systems, respectively. The ratio of $f_S$ and $f_M$ can be calculated in the following equation:
\begin{equation} 
\label{eq:fsfm}
\frac{f_S}{f_M}=\cfrac{\cfrac{N_S}{1-\rm{MR}}}{\cfrac{N_M}{\rm{MR}}}.
\end{equation}
With Equation \ref{eq:fre_all} and \ref{eq:fsfm}, $f_S$ and $f_M$ can be solved independently given that $N_S$ and $N_M$ can be measured and that $f$ can be measured globally\citep[e.g., ][]{Fressin2013}. In addition, the MR for planet host stars {{(${\rm{MR}}_{\rm{PL}}$)}} can be calculated and compared to a global MR:
\begin{equation} 
\label{eq:mfpl}
{\rm{MR}}_{\rm{PL}}=\cfrac{N_M}{N_M+N_S},
\end{equation}

\subsection{Stellar Multiplicity Rate For Planet Host Stars}
\begin{figure}[htp]
\begin{center}
\includegraphics[angle=0, width= 0.45\textwidth]{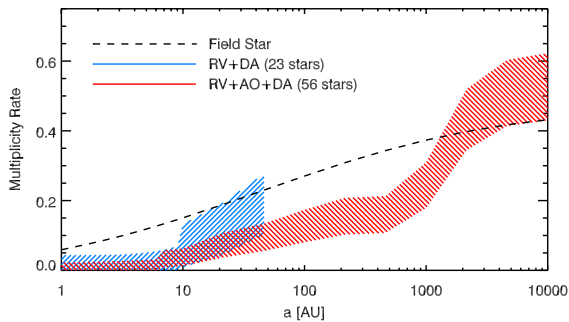} 
\caption{Comparison of stellar multiplicity rate between field stars (dashed line) and planet host stars (hatched regions). Blue hatched region represents the 1-$\sigma$ region of the stellar multiplicity rate for 23 planet host stars with RV and DA analysis~\citep{Wang2014}. AO data were not incorporated, so the sensitivity of RV and DA was limited within 50 AU. For this study, AO data are used to constrain stellar companions beyond 50 AU. Red hatched region represents the 1-$\sigma$ region of the stellar multiplicity rate for 56 stars with RV, AO and DA analysis. The new study shows that the stellar multiplicity rate for planet host stars is lower than that for the field stars within 1500 AU, indicating a more profound influence of stellar companions on planet formation.   
\label{fig:Multi_Field}}
\end{center}
\end{figure}

Fig. \ref{fig:Multi_Field} shows the comparison between the stellar multiplicity rate for field stars~\citep[dashed line][]{Duquennoy1991,Raghavan2010} and that for planet host stars (blue and red hatched regions). The red hatched region is the 1-$\sigma$ uncertainty region for 56 stars with RV and AO observations, and the DA analysis. The error bar of $N_M$ is estimated based on Poisson statistics. The square root of the closest integer to $N_M$ is used as the error bar to $N_M$ unless  the closest integer is zero, in which case we used 1 for the error of $N_M$. The stellar multiplicity rate for planet host stars is significantly lower than that of field stars until the separation reaches $\sim$1500 AU. This implies that the influence of a stellar companion may be more profound than previously thought. The effective separation below which planet formation is significantly affected is extended to $\sim$1500 AU. In comparison, the blue hatched region represents the 1-$\sigma$ uncertainty region for 23 stars with RV data and DA analysis, but no AO observations~\citep{Wang2014}. Based on the blue hatched region, the significant difference of stellar multiplicity disappears after separation reaches 20.8 AU. Since ~\citet{Wang2014}, we have incorporated AO data into our analyses and increased the sample size from 23 to 56. These improvements greatly strengthen the statistics in the comparison. Specifically, increasing the sample size reduces the statistical uncertainty; adding AO data helps constrain stellar companions beyond the reach of the RV technique. 

\subsection{Planet Occurrence Rate vs. Binary Separation}
\begin{figure}[htp]
\begin{center}
\includegraphics[angle=0, width= 0.45\textwidth]{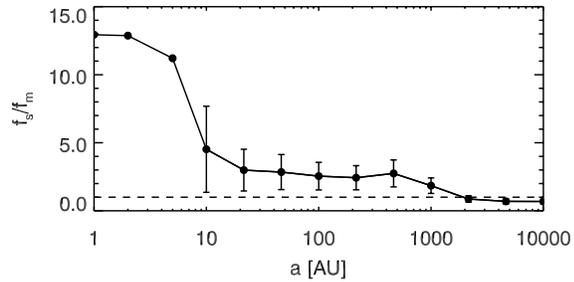} 
\caption{The ratio of the planet occurrence rates for single and multiple-stars. Dashed line represents the value of 1, a value indicating a comparable planet occurrence rate. The planet occurrence rate for single stars is much higher than that for multiple-stars within 10 AU. Beyond 10 AU, the ratios are $4.5\pm3.2$, $2.6\pm1.0$, and $1.7\pm0.5$ for 10 AU, 100 AU, and 1000 AU, respectively, indicating planets in multiple-star systems are fewer than those around single stars at these separations. The planet occurrence rates become comparable between single and multiple-stars when separation is larger than $\sim$1500 AU. Error bars are calculated based on Poisson statistics and propagated through Equation \ref{eq:fsfm}. No error bar is shown within 10 AU because of less than 1 detection of stellar companion according to Equation \ref{eq:nmns}. 
\label{fig:Multi_single}}
\end{center}
\end{figure}

With the stellar multiplicity rate for planet host stars, we can calculate the ratio of the planet occurrence rate for single and multiple-star systems according to Equation \ref{eq:fsfm}. Fig. \ref{fig:Multi_single} shows the ratio $f_S/f_M$ as a function separation. Planets orbiting single stars are $4.5\pm3.2$, $2.6\pm1.0$, and $1.7\pm0.5$ times more likely than planets in S-type orbits in multiple-star systems with stellar separations of 10 AU, 100 AU, and 1000 AU, respectively. The deficiency of planets around multiple-stars indicates that the suppressive influence on planet formation of a stellar companion is significant at these separations. The suppressive effect decreases as separation increases, and $f_S$ and $f_M$ are comparable at separations around $\sim$1500 AU, indicating that stellar companions at these separations barely have any influence on planet formation. The comparison of planet occurrence rate for single and multiple-stars at other stellar separations is given in Table \ref{tab:fsfm}.

\subsection{Comparison to Previous Results}
\label{sec:comp}
The field of studying planets in multiple-star systems may be divided into two eras: before and after the $Kepler$ mission. Before the $Kepler$ mission, stars with giant planets are the main targets, and they are mostly detected by the RV technique.~\citet{Bonavita2007} used a sample defined as the ``uniform detectability" (UD) sample. They searched for stellar companions around stars in this sample, and found that the fractions of stars with detected planets are comparable between single and multiple-stars. However, after considering the search incompleteness, they concluded that the frequency of planets in binary stars cannot be more than a factor of 3 lower than that for single stars. Their finding is consistent with our conclusion for separations larger than $\sim$50 AU. However, we find that $f_s/f_m$ can be higher than 3 for shorter binary separations (Table \ref{tab:fsfm}). ~\citet{Eggenberger2011} presented comparison of stellar multiplicity rate between planet host stars and a control sample of non-planet host stars. They concluded that S-type gas giant planets are less frequent in binary stars with mean semi-major axes between 35 and 250 AU. Their conclusion is qualitatively consistent with ours, but we find that planet formation can be suppressed at larger separations (out to 1500 AU). We emphasize that there are fundamental differences in the comparison to previous results on RV planet surveys. First, they focused on host stars of gas giant planets, whereas this study has made use of $Kepler$ data, and therefore mostly deals with lower-mass planets. Second, RV surveys have a much stronger bias against close-in binary stars than the $Kepler$ mission.

After the $Kepler$ satellite was launched, studies continue on the stellar multiplicity of planet host stars.~\citet{LilloBox2012} found that the visual companion rate for KOIs is 17.3\% and 41.8\% within 3$^{\prime\prime}$ and 6$^{\prime\prime}$, respectively. {{They later updated the companion rate to be 17.2\% and 32.8\% within 3$^{\prime\prime}$ and 6$^{\prime\prime}$~\citep{LilloBox2014}.~\citet{Dressing2014} found that  17.2\% of KOIs have visual companions within 3$^{\prime\prime}$. In comparison, we find that $12.5\pm4.7$\% and $48.2\pm9.3$\% of KOIs have visual companions within 3$^{\prime\prime}$ and 6$^{\prime\prime}$, which is consistent with their numbers.}}~\citet{Adams2012} found that 60\%, 20\%, and 7\% of 90 KOIs have stellar companions within 6$^{\prime\prime}$, 2$^{\prime\prime}$, and 0$^{\prime\prime}$.5, respectively. We find that these numbers to be $48.2\pm9.3$\%, $5.4\pm3.1$\%, and $3.6\pm2.5$\%. In comparison, we find significantly fewer stellar companions than~\citet{Adams2012} at angular separations between 0$^{\prime\prime}$.5 to 2$^{\prime\prime}$.0. 

We therefore conduct a cross check with their targets, and find 20 overlapping targets. For these targets, we detect 40 companions using the images from the CFOP, while they detect 33 companions. {{We find 17 new companions that were not reported by~\citet{Adams2012}. Most of the new companions are more than 6$^{\prime\prime}$ away from central stars. We are not able to detect 10 of their companions}}. All of our missing detections have $\Delta$ Mag larger than 7.1 mag (close to detection limit, see Fig. \ref{fig:AO_contrast}), and none of them are within 2$^{\prime\prime}$ except for KOI 18 (0$^{\prime\prime}$.9 separation and $\Delta$ mag = 5.0). We suspect the difference may be a result of different thresholds for companion detections or differences in manual inspections. 

We also conduct investigations on the lack of companion detections within 2$^{\prime\prime}$.0. In the overlapping sample of 20 KOIs with~\citet{Adams2012}, we detect 2 companions within 0$^{\prime\prime}$.5, KOI 292 (0$^{\prime\prime}$.43), and KOI 975 (0$^{\prime\prime}$.72). They are also detected in~\citet{Adams2012}, but KOI 18 with a separation of 0$^{\prime\prime}$.9 was missed in our search. For the overlapping sample, $10.0\pm7.1$\% (2 out 20) have companions within 2$^{\prime\prime}$. In comparison, for the rest of our sample, none of 36 stars has detected companions within 2$^{\prime\prime}$, which raises a concern that KOIs with close-in companions may be filtered out when conducting RV followup observations. However, it does not seem to be the case for KOI 18, KOI 292, and KOI 975, these targets receive continued RV followup observations even after close-in companions are detected in AO images. 

\section{Summary and Discussion}
\label{sec:Summary}

\subsection{Summary}
\label{sec:sum}

We conduct a search for stellar companions to a sample of 56 $Kepler$ planet host stars, and compare the stellar multiplicity rate for planet host stars and the field stars in the solar neighborhood. We find that the stellar multiplicity rate for planet host stars is significantly lower than that for the field stars at stellar separations smaller than 1500 AU, indicating that planet formation is less efficient in multiple-star systems than in single stars. The influence of stellar companions plays a significant role in planet formation and evolution in multiple-star systems with separations smaller than 1500 AU. 

We distinguish the planet occurrence rates for single and multiple-stars. We find that planets in S-type orbits in multiple-star systems are $4.5\pm3.2$, $2.6\pm1.0$, and $1.7\pm0.5$ times less frequent than planets orbiting single stars if a stellar companion is present at distances of 10, 100 and 1000 AU, respectively. The difference in planet occurrence rate between single and multiple-star systems becomes insignificant when companions separation exceeds 1500 AU, suggesting that planet formation in {{widely-separated binaries}} is similar to that around single stars. 

In summary, three improvements in this study allow us to better study planets in multiple-star systems. First, unlike planet host stars selected from ground-based RV and transiting surveys, our sample from the $Kepler$ mission does not have strong bias against planets in multiple-star systems. Second, we combine the RV and AO data for the 56 $Kepler$ stars, which construct a survey for stellar companions with high completeness. The DA method is also used to put further constraints on stellar companions in systems with multiple transiting planets. Third, we develop a method to quantify the detection bias of planets in multiple-star systems, which enables a fair comparison of stellar multiplicity rate.

\subsection{Discussion}
\label{sec:dis}

\subsubsection{Stellar Companions to Hot Jupiter Host Stars}
There are 6 hot Jupiter (HJ, $P<10$ day and R$_P > 5$ R$_\oplus$) host stars in our sample. They are KOI 5, KOI 10, KOI 17, KOI 18, KOI 20, and KOI 22. Four (67\%) of them have detected stellar companions. The stellar multiplicity rate for HJ host stars is much higher than the rest of the sample, i.e., 32\%. While we recognize the small number statistics and the possible non-association of these visual companions, this may imply that stellar companions play a role in HJ migration.~\citet{Knutson2014} conducted a search for massive companions to close-in gas giant planets. They estimated an occurrence rate of 51\%$\pm$10\% for companions with masses between 1-13 M$_J$ and semi-major axes between 1-20 AU. The high occurrence rate for  both massive sub-stellar companions and stellar companions may suggest that planet-planet and star-planet interactions have a comparable influence on the migration of HJs. Given the large separations of stellar companions($a>1500$ AU), the Kozai timescales for all HJ systems with stellar companions (except for KOI 5) are $\sim10^8$-$10^9$ years, which are comparable to the age of the systems, and perhaps too long to effectively perturb the orbit of a gas giant planet. Therefore, it is still inconclusive whether the HJs in these systems migrate to their current positions due the perturbation of the detected stellar companions. 

\subsubsection{Stellar Multiplicity Rate For Single and Multiple Planet Systems}
\begin{figure}[htp]
\begin{center}
\includegraphics[angle=0, width= 0.45\textwidth]{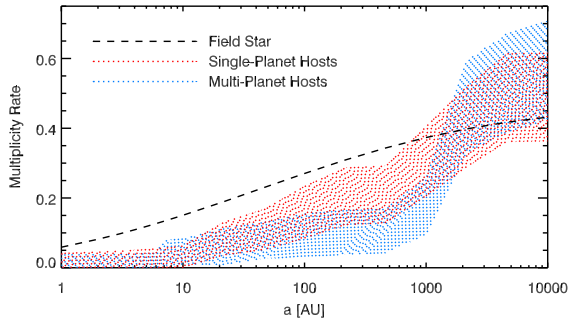} 
\caption{Comparison of stellar multiplicity rate for field stars (dashed line), 29 planet host stars with a single detected planet (red dotted region, 1-$\sigma$ range), 27 planet host stars with multiple detected planets (blue dotted region, 1-$\sigma$ range). 
\label{fig:Multi_single_field}}
\end{center}
\end{figure}

Perturbation from a companion star will change the mutual inclinations of planets in the same system~\citep[][see also \S \ref{sec:dynamical}]{Wang2014}. We therefore expect to see a lower stellar multiplicity rate for stars with multiple transiting planets than stars with only one transiting planet. There are 27 stars in our sample with multiple transiting planets and 29 stars with only one transiting planet. Fig. \ref{fig:Multi_single_field} shows the comparison of stellar multiplicity rate for these two sub samples. The hatched regions with different colors overlap, so there is no statistically significant difference in the stellar multiplicity rate between systems with multiple transiting planets and system with only one transiting planet. However, for separations between 50 and 1000 AU, we notice a relatively lower stellar multiplicity rate for multiple transiting planet systems than systems with only one detected transiting planet, suggesting that companion perturbations affect planet mutual inclination and/or multiple planet formation. An ongoing AO campaign is being carried out at the Palomar observatory to study the stellar multiplicity rate for multi-planet host stars, and will address the role of stellar perturbation in planet formation and detection. 

\subsubsection{Needing a Better Control Sample Than the Field Stars}
There are several uncertainties to use the field stars as a control sample to compare to the $Kepler$ sample. First, it is uncertain that $Kepler$ overall sample (i.e., all $Kepler$ stars) multiplicity rate is the same as the field stars. $Kepler$ stars are mainly selected by applying a magnitude cut (magnitude-limited), whereas the field stars are volume-limited~\citep{Duquennoy1991, Raghavan2010}. Therefore, Malmquist bias exsits for the $Kepler$ stars, brighter, more distant stars, are selected, which include young stars, giant stars, and binary stars. While some young stars and giant stars can be excluded by applying a T$_{\rm{eff}}$ and $log(g)$ cut from the $Kepler$ Input Catalog~\citep[KIC,][]{Brown2011}, it is more difficult to discern binary stars, so it is possible that the stellar multiplicity rate for $Kepler$ stars could be higher than that of the field stars~\citep{Gaidos2013b}. 

The second uncertainty lies in the fraction of field stars with planets. We compare the stellar multiplicity rate for the field stars and planet host stars. Most of them are small planet host stars. However, {{the fraction of field stars hosting small planets is less constrained than the fraction of field stars hosting large planets~\citep{Howard2010}}}. If (1) not all field stars have a planet; and (2) the statistics of multiple-stars (multiplicity, separation distribution, etc.) are comparable for the nearby solar-type stars and for the stars in our sample, then the difference in Fig. \ref{fig:Multi_Field} should suggest the impact of stellar companions on planet occurrence. In this case, the field stars are a sample contaminated by planet host stars. If we see a difference when comparing a sample of planet host stars to the field stars, then the difference would have been more distinct when comparing the planet host sample and a non-planet host sample. The latter is difficult to obtain because of the limitation of current detection sensitivity. However, a planet mass or radius limit can be set to study a certain type of planet, e.g., comparing the stellar multiplicity rate for the giant planet host stars and stars without a gas giant planet. In this case, any difference in the multiplicity rate reflects the impact of stellar companions on gas giant planet formation. 

\noindent{\it Acknowledgements} The authors thank Howard Isaacson and Matt Giguere for helpful comments and proofreading the paper. The research is made possible by the data from the Kepler Community Follow-up Observing Program (CFOP). The authors acknowledge all the CFOP users who uploaded the AO and RV data used in the paper. This research has made use of the NASA Exoplanet Archive, which is operated by the California Institute of Technology, under contract with the National Aeronautics and Space Administration under the Exoplanet Exploration Program. Jiwei Xie is supported by a Foundation for the Author of National Excellent Doctoral Dissertation (FANEDD) of PR China.

\clearpage

%%%%%%%%%%%%%%%%%%%%%%%%%%%%%%%%%%%%%%%%%
%			TABLES
%%%%%%%%%%%%%%%%%%%%%%%%%%%%%%%%%%%%%%%%%

\newpage
\LongTables
%\begin{landscape}
\begin{deluxetable*}{lccccccccccll}
\LongTables%\tablewidth{0pt}
%\tabletypesize{\tiny}
\setlength{\tabcolsep}{0.02in}
%\tablenum{1}
\tablecaption{RV and AO data for 56 KOIs\label{tab:stellar_params}}
\tablehead{
%\colhead{\textbf{ }} &
%\multicolumn{5}{c}{\textbf{KOI}} &
%\multicolumn{7}{c}{\textbf{Derived Values}} \\
\multicolumn{8}{c}{\textbf{KOI}} &
\multicolumn{3}{c}{\textbf{RV}} &
\multicolumn{2}{c}{\textbf{AO}} \\
\colhead{\textbf{KOI}} &
\colhead{\textbf{KIC}} &
\colhead{\textbf{$\alpha$}} &
\colhead{\textbf{$\delta$}} &
\colhead{\textbf{K$_P$}} &
\colhead{\textbf{$T_{\rm eff}$}} &
\colhead{\textbf{$\log g$}} &
\colhead{\textbf{\#PL}} &
\colhead{\textbf{$T_{\rm start}$}} &
\colhead{\textbf{$T_{\rm end}$}} &
\colhead{\textbf{\#RV}} &
\colhead{\textbf{Telescope}} &
\colhead{\textbf{Band}} \\
\colhead{\textbf{}} &
\colhead{\textbf{}} &
\colhead{\textbf{(deg)}} &
\colhead{\textbf{(deg)}} &
\colhead{\textbf{(mag)}} &
\colhead{\textbf{(K)}} &
\colhead{\textbf{(cgs)}} &
\colhead{\textbf{}} &
\colhead{\textbf{(MJD)}} &
\colhead{\textbf{(MJD)}} &
\colhead{\textbf{}} &
\colhead{\textbf{}} &
\colhead{\textbf{}} 
}

\startdata

00005&8554498&289.739716&44.647419&11.665&5861.00&4.190&2& 54983.516& 56486.440&21&Keck Palomar&Br-$\gamma$ J \\
00007&11853905&285.615326&50.135750&12.211&5858.00&4.280&1& 55041.491& 56134.480&22&Keck Palomar&H J \\
00010&6922244&281.288116&42.451080&13.563&6025.00&4.110&1& 54983.540& 55781.534&50&Palomar&J \\
00017&10874614&296.837250&48.239944&13.303&5826.00&4.420&1& 54984.561& 55043.520&10&Palomar&J \\
00018&8191672&299.407013&44.035053&13.369&6297.00&3.990&1& 54985.594& 55110.314&9&Gemini Gemini Palomar&R Y J \\
00020&11804465&286.243439&50.040379&13.438&6011.00&4.230&1& 55014.412& 55761.325&16&Palomar&J \\
00022&9631995&282.629669&46.323360&13.435&5972.00&4.410&1& 55014.403& 55792.438&16&Palomar&J \\
00041&6521045&291.385986&41.990269&11.197&5909.00&4.300&3& 54988.511& 56533.359&64&Keck Palomar&Br-$\gamma$ J \\
00069&3544595&291.418304&38.672359&9.931&5593.00&4.510&1& 55042.587& 56547.339&61&Keck MMT MMT Palomar&Br-$\gamma$ J K J \\
00070&6850504&287.697998&42.338718&12.498&5443.00&4.450&5& 55073.386& 56533.484&38&Palomar&J \\
00072&11904151&285.679382&50.241299&10.961&5627.00&4.390&2& 55073.400& 56172.301&54&MMT MMT Palomar&J K J \\
00082&10187017&281.482727&47.208031&11.492&4908.00&4.610&5& 55311.579& 56533.335&65&Keck MMT MMT&Br-$\gamma$ J K \\
00084&2571238&290.420807&37.851799&11.898&5541.00&4.530&1& 55073.419& 55723.450&20&Palomar&J \\
00085&5866724&288.688690&41.151180&11.018&6172.00&4.360&3& 55696.490& 55738.516&6&MMT MMT&J K \\
00087&10593626&289.217499&47.884460&11.664&5510.00&4.500&1& 55425.386& 56521.486&26&Palomar Palomar&J K \\
00103&2444412&291.683350&37.751591&12.593&5531.00&4.440&1& 55073.441& 55797.539&16&Palomar&J \\
00104&10318874&281.194733&47.497131&12.895&4786.00&4.590&1& 55377.342& 56525.316&30&Keck Palomar Palomar&K J K \\
00108&4914423&288.984558&40.064529&12.287&5975.00&4.330&2& 55073.469& 56145.498&22&Keck Palomar&K J \\
00111&6678383&287.604614&42.166779&12.596&5711.00&4.410&4& 55372.555& 56521.442&14&Palomar&J \\
00116&8395660&300.863983&44.337551&12.882&5865.00&4.410&4& 55133.397& 56532.332&33&Keck MMT&K K \\
00122&8349582&284.482452&44.398041&12.346&5714.00&4.390&1& 55073.495& 56151.511&33&Keck Palomar&K J \\
00123&5094751&290.392731&40.284870&12.365&5871.00&4.150&2& 55074.492& 56166.345&15&Keck Palomar&K J \\
00137&8644288&298.079437&44.746319&13.549&5385.00&4.430&3& 55075.509& 56146.484&20&Palomar&J \\
00148&5735762&299.139221&40.949020&13.040&5190.00&4.490&3& 55073.527& 56532.436&43&Keck Palomar&K J \\
00153&12252424&287.997894&50.944328&13.461&4725.00&4.640&2& 55313.592& 56524.353&29&Keck MMT Palomar Palomar&Br-$\gamma$ K J K \\
00157&6541920&297.115112&41.909142&13.709&5685.00&4.380&6& 55440.501& 56533.434&7&Palomar Robo-AO&LP600 \\
00180&9573539&284.394287&46.249081&13.024&5680.00&4.500&1& 55074.466& 55083.341&6&Palomar&J \\
00244&4349452&286.638397&39.487881&10.734&6103.00&4.070&2& 55366.603& 56519.408&104&Keck Palomar Palomar&Br-$\gamma$ J K \\
00245&8478994&284.059540&44.518215&9.705&5288.00&4.590&4& 55312.586& 56523.237&59&Gemini Gemini Keck MMT MMT Palomar&R Y Br-$\gamma$ J K K \\
00246&11295426&291.032318&49.040272&9.997&5793.00&4.281&2& 55312.582& 56519.420&65&Keck MMT MMT&Br-$\gamma$ J K \\
00261&5383248&297.069611&40.525131&10.297&5692.00&4.420&1& 55404.530& 56518.546&36&Keck MMT MMT&Br-$\gamma$ J K \\
00263&10514430&281.273804&47.774399&10.821&5820.00&4.150&1& 55395.529& 55788.482&6&MMT MMT Palomar Palomar&J K J K \\
00265&12024120&297.018829&50.408981&11.994&6040.00&4.360&1& 55782.522& 56532.354&27&Gemini Gemini Palomar Palomar&R Y J K \\
00273&3102384&287.478516&38.228840&11.457&5783.00&4.430&1& 55431.309& 56613.223&20&MMT MMT&J K \\
00274&8077137&282.492218&43.980209&11.390&6090.00&4.130&2& 55403.447& 56474.522&8&Gemini Gemini MMT MMT&R Y J K \\
00283&5695396&288.530823&40.942299&11.525&5687.00&4.420&2& 55433.368& 56524.331&30&Keck Palomar Palomar&Br-$\gamma$ J K \\
00289&10386922&282.945648&47.574905&12.747&5812.00&4.458&2& 56449.401& 56532.291&5&Lick&J \\
00292&11075737&287.326630&48.673431&12.872&5780.00&4.430&1& 55377.475& 56166.411&21&Keck Palomar Palomar&K J K \\
00299&2692377&285.661652&37.964500&12.899&5538.00&4.340&1& 55403.512& 56530.414&32&Keck&K \\
00305&6063220&297.354004&41.300049&12.970&4782.00&4.610&1& 55403.543& 56531.340&36&Keck&K \\
00321&8753657&291.848053&44.968220&12.520&5538.00&4.340&2& 55378.534& 56493.368&47&Keck Lick&K J \\
00364&7296438&295.872314&42.881149&10.087&5798.00&4.240&1& 55376.346& 55699.442&3&WIYN&$i$ \\
00365&11623629&297.486908&49.623451&11.195&5451.00&4.490&2& 55402.354& 56532.377&24&Palomar Palomar&J K \\
00377&3323887&285.573975&38.400902&13.803&5777.00&4.450&3& 55342.448& 56506.363&16&Palomar Palomar&J K \\
00701&9002278&283.212738&45.349861&13.725&4807.00&4.690&5& 56137.475& 56507.530&15&Keck&K \\
00975&3632418&287.361816&38.714016&8.224&6131.00&3.950&1& 55439.438& 56486.562&44&MMT MMT&J K \\
01431&11075279&287.022278&48.681938&13.460&5649.00&4.460&1& 56472.391& 56532.502&6&WIYN&$i$ \\
01439&11027624&290.851776&48.521339&12.849&5930.00&4.090&1& 55075.273& 56531.312&6&Lick&J \\
01442&11600889&286.036346&49.614510&12.521&5476.00&4.448&1& 55696.518& 56446.424&17&Keck Lick&K J \\
01463&7672940&288.258636&43.376465&12.328&6020.00&4.380&1& 56027.511& 56530.292&3&WIYN&$i$ \\
01612&10963065&284.786194&48.423229&8.769&6027.00&4.220&1& 55697.588& 56494.420&42&Keck Lick&Br-$\gamma$ J \\
01781&11551692&287.605591&49.523258&12.231&4977.00&4.590&3& 56076.566& 56112.331&9&Palomar Robo-AO&LP600 \\
01925&9955598&293.679199&46.852760&9.439&5365.00&4.430&1& 55999.646& 56547.332&40&Keck Palomar&Br-$\gamma$ K \\
02169&9006186&285.207489&45.384350&12.404&5447.00&4.420&4& 56099.455& 56171.456&4&Palomar Robo-AO&LP600 \\
02687&7202957&292.615112&42.764252&10.158&5803.00&3.910&2& 55999.652& 56531.383&22&Palomar&K \\
02720&8176564&295.439667&44.039162&10.338&6109.00&4.410&1& 56018.575& 56519.389&18&Keck Palomar&Br-$\gamma$ K \\

\enddata

%\tablecomments{Columns marked with a $^{\dagger}$ are values from NASA Exoplanet Archive. Stars marked with a $^{\dagger\dagger}$ have missing $T_{\rm{eff}}$. Their $T_{\rm{eff}}$ values are estimated based on $g-r$ and $g-i$ values~\citep{Pinsonneault2012}, the adopted $T_{\rm{eff}}$ is the average of the results. There are three exceptions, values of $T_{\rm{eff}}$ for K03158 and K0368 are adopted from~\citet{Ammons2006} and $T_{\rm{eff}}$ for K04021 is based on the infrared flux method in ~\citet{Pinsonneault2012}}

\end{deluxetable*}
%\end{landscape}
% \global\pdfpageattr\expandafter{\the\pdfpageattr/Rotate 90}
%\input{stellar_params.tex} % stellar parameters of planet hosts.
%\newpage
%\input{orbital_params.tex} % stellar parameters of planet hosts.
%\newpage
%\input{AO_params.tex} % stellar parameters of planet hosts.

\begin{deluxetable*}{lcccccccc}
\tablewidth{0pt}
%\tabletypesize{\tiny}
%\setlength{\tabcolsep}{0.02in}
\tablecaption{RV measurement results for 56 KOIs.\label{tab:rv}}
\tablehead{
\colhead{{KOI}} &
\colhead{{KIC}} &
\colhead{{RMS$_1^{\ast\ast}$}} &
\colhead{{$\delta v$}} &
\colhead{{$\frac{\rm{RMS}_1}{\delta v}>5$}} &
\colhead{{Slope}} &
\colhead{{Non-transiting}} &
\colhead{{RMS$_2^{\ast\ast\ast}$}} &
\colhead{{$\frac{\rm{RMS}_2}{\delta v}>5$}} \\
\colhead{{}} &
\colhead{{}} &
\colhead{{(m$\cdot$s$^{-1}$)}} &
\colhead{{(m$\cdot$s$^{-1}$)}} &
\colhead{{}} &
\colhead{{}} &
\colhead{{}} &
\colhead{{(m$\cdot$s$^{-1}$)}} &
\colhead{{}} \\
}

\startdata

00005$^\ast$&8554498& 94.9& 4.2&$\surd$& &$\surd$& 23.7&$\surd$ \\
00007&11853905& 7.9& 2.1& & & & 4.5& \\
00010&6922244& 53.8& 13.2& & & & 47.1& \\
00017&10874614& 57.8& 3.6&$\surd$& & & 7.1& \\
00018&8191672& 178.1& 6.1&$\surd$& & & 18.2& \\
00020&11804465& 44.5& 4.1&$\surd$& & & 22.0&$\surd$ \\
00022&9631995& 44.2& 3.8&$\surd$& & & 40.9&$\surd$ \\
00041&6521045& 5.4& 1.6& & & & 5.4& \\
00069$^\ast$&3544595& 21.2& 1.3&$\surd$&$\surd$& & 3.0& \\
00070&6850504& 11.0& 1.9&$\surd$& & & 10.2&$\surd$ \\
00072&11904151& 4.6& 1.6& & & & 3.7& \\
00082&10187017& 4.5& 1.2& & & & 4.2& \\
00084&2571238& 11.2& 1.8&$\surd$& & & 11.2&$\surd$ \\
00085&5866724& 10.1& 1.9&$\surd$& & & 15.5&$\surd$ \\
00087&10593626& 5.0& 1.6& & & & 4.8& \\
00103&2444412& 6.8& 1.9& & & & 6.2& \\
00104&10318874& 96.1& 2.0&$\surd$& &$\surd$& 6.1& \\
00108&4914423& 7.6& 2.2& & & & 7.4& \\
00111&6678383& 6.1& 2.6& & & & 6.6& \\
00116&8395660& 6.5& 2.4& & & & 6.5& \\
00122&8349582& 5.1& 1.7& & & & 4.7& \\
00123&5094751& 7.1& 2.4& & & & 6.9& \\
00137&8644288& 9.7& 2.5& & & & 8.0& \\
00148&5735762& 32.7& 2.3&$\surd$& &$\surd$& 14.2&$\surd$ \\
00153&12252424& 9.7& 2.4& & & & 8.9& \\
00157&6541920& 36.7& 8.1& & & & 49.9&$\surd$ \\
00180&9573539& 15.7& 2.1&$\surd$& & & 15.6&$\surd$ \\
00244&4349452& 8.3& 3.5& & &$\surd$& 6.5& \\
00245&8478994& 5.0& 1.3& & & & 3.3& \\
00246&11295426& 16.9& 1.3&$\surd$& &$\surd$& 3.6& \\
00261&5383248& 5.2& 1.5& & & & 5.3& \\
00263&10514430& 11.1& 3.2& & & & 13.9& \\
00265&12024120& 4.7& 1.8& & & & 4.5& \\
00273&3102384& 91.7& 1.4&$\surd$& & & 10.3&$\surd$ \\
00274&8077137& 4.0& 1.9& & & & 3.7& \\
00283&5695396& 5.7& 1.6& & & & 5.7& \\
00289&10386922& 12.1& 2.1&$\surd$& & & 12.1&$\surd$ \\
00292&11075737& 6.2& 2.3& & & & 4.5& \\
00299&2692377& 6.6& 1.9& & & & 6.0& \\
00305&6063220& 5.8& 1.7& & & & 5.6& \\
00321&8753657& 4.3& 1.7& & & & 4.0& \\
00364&7296438& 51.3& 0.3&$\surd$& & & 51.3&$\surd$ \\
00365&11623629& 4.3& 1.3& & & & 4.0& \\
00377&3323887& 13.3& 6.0& & & & 12.3& \\
00701&9002278& 4.7& 3.0& & & & 4.9& \\
00975&3632418& 8.9& 3.0& & & & 8.9& \\
01431&11075279& 8.3& 2.1& & & & 7.1& \\
01439&11027624& 10.2& 2.3& & & & 15.4&$\surd$ \\
01442&11600889& 89.0& 2.0&$\surd$& &$\surd$ & 3.1& \\
01463&7672940& 116.1& 2.8&$\surd$& & & 116.1&$\surd$ \\
01612&10963065& 4.1& 1.5& & & & 4.0& \\
01781&11551692& 21.7& 1.4&$\surd$& & & 24.3&$\surd$ \\
01925&9955598& 3.4& 1.1& & & & 2.5& \\
02169&9006186& 6.6& 0.9&$\surd$& & & 6.6&$\surd$ \\
02687&7202957& 9.2& 1.3&$\surd$& & & 10.2&$\surd$ \\
02720&8176564& 3.8& 1.4& & & & 3.6& \\

\enddata

\tablecomments{$\ast$: KOIs considered in multiple-star systems. See \S \ref{sec:koi05} and \S \ref{sec:koi69} for detailed discussions. $\ast\ast$: RMS of the RV measurements. $\ast\ast\ast$: RMS after removing the linear trend or the long-period signal, and the RV signal caused by detected planet candidates. }

\end{deluxetable*}

%\input{AO_params.tex} % stellar parameters of planet hosts.

%\LongTables
\begin{deluxetable*}{lcccccc}
%\tabletypesize{\tiny}
\tablewidth{0pt}
%\tablenum{1}
\tablecaption{Visual companion detections with AO data.\label{tab:AO_params}}
\tablehead{
\colhead{\textbf{KOI}} &
\colhead{\textbf{$\Delta$ Mag}} &
%\colhead{\textbf{Separation}} &
\multicolumn{2}{c}{\textbf{Separation}} &
%\colhead{\textbf{Separation}} &
\colhead{\textbf{Distance}} &
\colhead{\textbf{Significance}} &
\colhead{\textbf{PA}} \\
\colhead{\textbf{}} &
\colhead{\textbf{(mag)}} &
\colhead{\textbf{(arcsec)}} &
\colhead{\textbf{(AU)}} &
\colhead{\textbf{(pc)}} &
\colhead{\textbf{($\sigma$)}} &
\colhead{\textbf{(deg)}} \\
}

\startdata

%from /home/jwang/2014/03/JW_2013_plot/ContrastCurve
   K00005 &   2.3 (Br-$\gamma$)&      0.14&      40.9&$  290.9^{+   63.2}_{  -19.4}$&      28.2&            308.9 \\
  K00010 &           6.8 ($J$)&      3.88&    3663.9&$  944.5^{+  100.1}_{ -139.1}$&      22.0&             89.3 \\
  K00017 &           3.9 ($J$)&      4.11&    2130.7&$  517.9^{+   27.6}_{  -28.4}$&     206.2&             39.5 \\
  K00018 &           3.9 ($J$)&      7.26&    8241.0&$ 1135.9^{+   84.4}_{ -154.0}$&     323.2&            148.2 \\
  K00018 &           6.3 ($J$)&      9.68&   10995.5&$ 1135.9^{+   84.4}_{ -154.0}$&      32.1&            344.7 \\
  K00018 &           6.6 ($J$)&      3.50&    3971.8&$ 1135.9^{+   84.4}_{ -154.0}$&      28.5&            110.1 \\
  K00018 &           7.3 ($J$)&      5.09&    5783.0&$ 1135.9^{+   84.4}_{ -154.0}$&      14.1&            211.3 \\
  K00018 &           7.7 ($J$)&      5.89&    6693.3&$ 1135.9^{+   84.4}_{ -154.0}$&       9.8&            106.3 \\
  K00018 &           7.7 ($J$)&     10.82&   12293.3&$ 1135.9^{+   84.4}_{ -154.0}$&       7.7&            222.1 \\
  K00018 &           7.3 ($J$)&      7.26&    8241.0&$ 1135.9^{+   84.4}_{ -154.0}$&       7.3&              77.6\\
  K00018 &           8.0 ($J$)&      9.69&   11004.0&$ 1135.9^{+   84.4}_{ -154.0}$&       6.6&            339.2 \\
  K00018 &           8.2 ($J$)&      7.09&    8059.2&$ 1135.9^{+   84.4}_{ -154.0}$&       6.5&            219.2 \\
  K00070 &           4.3 ($J$)&      3.79&    1058.9&$  279.5^{+   25.3}_{  -23.6}$&     217.7&             51.8 \\
  K00087 & 6.2 ($J$), 6.1 ($K$)&      5.49&     956.9&$  174.4^{+   15.2}_{  -12.1}$&      78.2&            177.2 \\
  K00087 & 7.4 ($J$), 6.6 ($K$)&      5.53&     964.1&$  174.4^{+   15.2}_{  -12.1}$&      28.4&             75.2 \\
  K00103 &           7.3 ($J$)&      9.81&    2985.6&$  304.4^{+   29.0}_{  -27.0}$&       9.5&            278.5 \\
  K00108 &           5.5 ($J$)&      9.52&    3357.4&$  352.7^{+   36.0}_{  -22.0}$&      98.8&            348.6 \\
  K00108 &           7.2 ($J$)&      5.00&    1764.4&$  352.7^{+   36.0}_{  -22.0}$&      21.5&            112.5 \\
  K00108 &           7.2 ($J$)&      2.51&     887.0&$  352.7^{+   36.0}_{  -22.0}$&      19.9&             74.8 \\
  K00108 &           7.4 ($J$)&      3.23&    1139.2&$  352.7^{+   36.0}_{  -22.0}$&      18.2&            100.9 \\
  K00108 &           7.4 ($J$)&      8.90&    3139.4&$  352.7^{+   36.0}_{  -22.0}$&      17.1&             19.2 \\
  K00111 &           7.5 ($J$)&      7.13&    2052.6&$  297.8^{+   26.0}_{  -29.1}$&      10.8&            117.7 \\
  K00111 &           7.8 ($J$)&      9.07&    2702.5&$  297.8^{+   26.0}_{  -29.1}$&       7.7&            175.5 \\
  K00111 &           8.2 ($J$)&      6.70&    1995.1&$  297.8^{+   26.0}_{  -29.1}$&       8.2&             96.0 \\
  K00116 &           3.8 ($K$)&      8.00&    2907.0&$  363.2^{+   56.7}_{  -40.3}$&     164.1&            353.5 \\
  K00116 &           4.8 ($K$)&     12.96&    4707.6&$  363.2^{+   56.7}_{  -40.3}$&      51.9&            144.3 \\
  K00116 &           6.2 ($K$)&      7.46&    2710.6&$  363.2^{+   56.7}_{  -40.3}$&      18.1&            107.1 \\
  K00116 &           6.3 ($K$)&     13.51&    4907.8&$  363.2^{+   56.7}_{  -40.3}$&      13.2&            113.8 \\
  K00116 &           6.3 ($K$)&     13.05&    4740.8&$  363.2^{+   56.7}_{  -40.3}$&      12.7&            357.4 \\
  K00116 &           7.5 ($K$)&     10.93&    3969.3&$  363.2^{+   56.7}_{  -40.3}$&       7.5&              19.7\\
  K00116 &           7.3 ($K$)&      5.79&    2101.4&$  363.2^{+   56.7}_{  -40.3}$&       6.3&            141.1 \\
  K00122 &           6.7 ($J$)&      4.23&    1446.2&$  341.7^{+   28.1}_{  -30.3}$&      30.0&            211.3 \\
  K00123 &           5.2 ($J$)&      9.52&    4749.5&$  498.7^{+   25.0}_{  -99.9}$&      62.3&            198.8 \\
  K00123 &           6.4 ($J$)&     10.19&    5083.7&$  498.7^{+   25.0}_{  -99.9}$&      19.2&             95.2 \\
  K00137 &           5.9 ($J$)&      5.64&    2471.4&$  438.2^{+   37.6}_{  -41.0}$&      44.4&            350.7 \\
  K00137 &           7.8 ($J$)&      7.13&    3122.2&$  438.2^{+   37.6}_{  -41.0}$&       7.6&            185.5 \\
  K00137 &           7.9 ($J$)&      5.11&    2240.3&$  438.2^{+   37.6}_{  -41.0}$&       6.6&            136.2 \\
  K00148 &           3.5 ($J$)&      4.44&    1369.1&$  308.7^{+   27.0}_{  -17.2}$&     519.8&            220.6 \\
  K00148 &           5.4 ($J$)&     10.99&    3391.9&$  308.7^{+   27.0}_{  -17.2}$&      77.0&            230.1 \\
  K00148 &           5.5 ($J$)&      2.54&     785.2&$  308.7^{+   27.0}_{  -17.2}$&      69.7&            245.8 \\
  K00148 &           6.3 ($J$)&      8.05&    2486.4&$  308.7^{+   27.0}_{  -17.2}$&      35.5&              244 \\
  K00148 &           7.4 ($J$)&      6.06&    1870.7&$  308.7^{+   27.0}_{  -17.2}$&      13.9&            238.8 \\
  K00153 &           6.0 ($K$)&      8.01&    1812.0&$  226.2^{+   18.6}_{  -15.1}$&      11.2&            353.4 \\
  K00153 & 6.9 ($J$), 7.6 ($K$)&      6.17&    1395.3&$  226.2^{+   18.6}_{  -15.1}$&       4.5&            298.4 \\
  K00244 & 2.7 ($J$), 2.0 ($K$)&      8.40&    2741.0&$  326.3^{+   23.4}_{  -44.6}$&    3231.4&            287.6 \\
  K00244 & 7.6 ($J$), 7.0 ($K$)&      8.38&    2733.6&$  326.3^{+   23.4}_{  -44.6}$&      27.5&            101.4 \\
  K00263 & 1.0 ($J$), 1.0 ($K$)&      3.28&     788.5&$  240.7^{+   17.7}_{  -38.9}$&    2430.8&            268.2 \\
  K00273 & 6.2 ($J$), 5.6 ($K$)&      5.02&    1201.8&$  239.6^{+   14.7}_{  -15.2}$&      32.6&            344.0 \\
  K00283 &           7.9 ($K$)&      6.09&    1266.2&$  208.0^{+   21.4}_{  -11.2}$&       7.0&           271.4  \\
  K00289 &           8.6 ($J$)&      5.86&    2201.2&$  375.9^{+  355.0}_{  -69.8}$&       5.1&             88.3 \\
  K00289 &           7.7 ($J$)&      3.18&    1195.0&$  375.9^{+  355.0}_{  -69.8}$&       6.1&            308.6 \\
  K00292 &           4.2 ($K$)&      0.43&     154.9&$  358.5^{+   40.9}_{  -28.3}$&      43.2&            119.4 \\
  K00365 & 7.7 ($J$), 6.6 ($K$)&      7.12&    1129.9&$  158.8^{+   15.1}_{  -19.7}$&      13.3&            313.7 \\
  K00377 & 5.0 ($J$), 4.8 ($K$) &      6.02&    3721.9&$  617.9^{+   48.5}_{  -46.7}$&     133.8&             91.9 \\
  K00377 & 6.0 ($J$), 6.9 ($K$)&      8.04&    4969.8&$  617.9^{+   48.5}_{  -46.7}$&      18.1&            221.9 \\
  K00377 & 6.2 ($J$), 7.3 ($K$)&      2.88&    1780.8&$  617.9^{+   48.5}_{  -46.7}$&      12.3&             37.5 \\
  K00975 & 3.8 ($J$), 4.0 ($K$)&      0.73&      90.4&$  123.7^{+    7.7}_{  -17.9}$&      31.2&            133.4 \\
  K01442 &           4.7 ($J$)&      2.11&     637.4&$  302.3^{+   18.0}_{  -20.3}$&      25.3&             76.3 \\
  K01442 &           8.0 ($J$)&      5.69&    1718.6&$  302.3^{+   18.0}_{  -20.3}$&       5.8&             90.2 \\

\enddata

\end{deluxetable*}

%\LongTables

\begin{deluxetable*}{lcccccccc}
%\tabletypesize{\tiny}
\tablewidth{0pt}
%\tablenum{1}
\tablecaption{Detection Bias of Planets in Multiple Stars.\label{tab:Bias_params}}
\tablehead{
\colhead{\textbf{KOI}} &
\colhead{\textbf{Period}} &
\colhead{\textbf{${\rm{R}}_P$}} &
\colhead{\textbf{${\rm{R}}_\ast$}} &
\colhead{\textbf{${\rm{M}}_\ast$}} &
\colhead{\textbf{Duration}} &
\colhead{\textbf{CDPP$_{\rm{eff}}^{\ast}$}} &
\colhead{\textbf{Quarters}} &
\colhead{\textbf{$\alpha^{\ast\ast}$}} \\
\colhead{\textbf{}} &
\colhead{\textbf{(day)}} &
\colhead{\textbf{(${\rm{R}}_\oplus$)}} &
\colhead{\textbf{(${\rm{R}}_\odot$)}} &
\colhead{\textbf{(${\rm{M}}_\odot$)}} &
\colhead{\textbf{(hr)}} &
\colhead{\textbf{(ppm)}} &
\colhead{\textbf{}} &
\colhead{\textbf{}} \\
}

\startdata

 K00005&	4.78033&   5.66&   1.42&   1.15&   2.01&      36.3&    17&  0.930\\
 K00007&	3.21366&   3.72&   1.27&   1.12&   4.11&      53.1&    14&  0.850\\
 K00010&	3.52250&  15.90&   1.56&   1.14&   3.20&     136.0&    17&  0.961\\
 K00017&	3.23470&  11.07&   1.08&   1.14&   3.60&     103.5&    14&  0.946\\
 K00018&	3.54847&  17.40&   2.02&   1.45&   4.08&     102.7&    17&  0.619\\
 K00020&	4.43796&  17.60&   1.38&   1.17&   4.67&     106.1&    14&  0.961\\
 K00022&	7.89145&  11.27&   1.11&   1.16&   3.79&      82.4&    17&  0.938\\
 K00041&       12.81570&   2.08&   1.23&   1.11&   6.54&      26.6&    17&  0.635\\
 K00069&	4.72675&   1.50&   0.87&   0.89&   2.93&      17.9&    17&  0.938\\
 K00070&       10.85410&   3.17&   0.94&   0.90&   3.82&      57.1&    17&  0.930\\
 K00072&	0.83749&   1.37&   1.00&   0.91&   1.80&      29.6&    14&  0.938\\
 K00082&       16.14570&   2.54&   0.74&   0.80&   3.75&      39.4&    17&  0.946\\
 K00084&	9.28701&   2.53&   0.86&   0.91&   3.54&      34.3&    17&  0.938\\
 K00085&	5.85993&   2.36&   1.20&   1.21&   4.11&      29.0&    17&  0.624\\
 K00087&      289.86200&   2.10&   0.85&   0.83&   7.40&      23.4&    17&  0.773\\
 K00103&       14.91080&   2.95&   0.95&   0.91&   3.31&      73.7&    17&  0.906\\
 K00104&	2.50806&   3.36&   0.76&   0.81&   1.14&      84.9&    17&  0.961\\
 K00108&       15.96530&   2.94&   1.21&   1.16&   4.65&      32.2&    17&  0.629\\
 K00111&       11.42750&   2.14&   0.93&   0.81&   4.59&      47.5&    17&  0.930\\
 K00116&       13.57070&   2.47&   1.04&   1.00&   3.25&      59.2&    17&  0.803\\
 K00122&       11.52310&   2.78&   1.09&   1.07&   4.06&      45.2&    17&  0.773\\
 K00123&	6.48167&   2.64&   1.43&   1.06&   3.63&      40.9&    17&  0.874\\
 K00137&       14.85890&   6.01&   0.98&   0.94&   3.63&      83.8&    17&  0.954\\
 K00148&	9.67393&   3.15&   0.89&   0.88&   4.40&      72.8&    17&  0.930\\
 K00153&	8.92511&   2.47&   0.69&   0.74&   2.77&      91.0&    17&  0.938\\
 K00157&       31.99550&   4.18&   1.06&   0.98&   4.27&      77.1&    17&  0.906\\
 K00180&       10.04560&   2.53&   0.92&   0.99&   3.26&      62.0&    17&  0.866\\
 K00244&       12.72040&   6.51&   1.66&   1.19&   2.83&      28.4&    17&  0.898\\
 K00245&       39.79220&   1.94&   0.73&   0.75&   4.57&      17.3&    17&  0.946\\
 K00246&	5.39877&   2.53&   1.24&   1.07&   3.56&      22.0&    17&  0.922\\
 K00261&       16.23850&   2.65&   1.02&   0.99&   3.86&      36.0&    17&  0.914\\
 K00263&       20.71940&   2.02&   1.41&   1.01&   4.23&      49.2&    17&  0.658\\
 K00265&	3.56806&   1.29&   1.18&   1.16&   3.43&      36.1&    17&  0.528\\
 K00273&       10.57380&   1.82&   1.07&   1.12&   1.74&      30.4&    17&  0.624\\
 K00274&       15.09200&   1.13&   1.55&   1.20&   4.14&      28.0&    17&  0.500\\
 K00283&       16.09190&   2.41&   1.03&   1.02&   2.93&      31.4&    17&  0.874\\
 K00289&      296.63700&   5.04&   0.95&   0.94&  16.43&      28.3&    17&  0.930\\
 K00292&	2.58663&   1.64&   0.98&   0.94&   2.37&      53.9&    14&  0.874\\
 K00299&	1.54168&   1.98&   1.11&   0.99&   1.94&      84.7&    17&  0.890\\
 K00305&	4.60356&   1.57&   0.73&   0.79&   2.40&      74.9&    17&  0.898\\
 K00321&	2.42631&   1.50&   1.11&   0.99&   2.65&      50.7&    17&  0.850\\
 K00364&      173.92800&   0.93&   1.35&   1.15&   2.64&      23.0&    17&  0.500\\
 K00365&       81.73750&   2.29&   0.87&   0.85&   6.78&      23.7&    17&  0.906\\
 K00377&       19.27390&   8.28&   1.01&   1.05&   4.16&     129.3&    17&  0.930\\
 K00701&       18.16410&   1.91&   0.60&   0.65&   2.96&      83.3&    17&  0.922\\
 K00975&	2.78582&   1.72&   2.04&   1.36&   3.41&      24.1&    17&  0.500\\
 K01431&      345.16100&   8.45&   1.00&   1.06&   7.50&      45.8&    14&  0.890\\
 K01439&      394.61100&   7.80&   1.65&   1.23&  24.61&      15.3&    17&  0.635\\
 K01442&	0.66934&   1.23&   0.99&   1.01&   1.29&      54.7&    14&  0.795\\
 K01463&      580.00000&  16.29&   1.09&   1.05&  11.43&      38.6&    17&  0.961\\
 K01612&	2.46503&   0.78&   1.31&   1.05&   1.19&      16.7&    14&  0.629\\
 K01781&	7.83445&   3.76&   0.76&   0.82&   3.00&      80.5&    14&  0.946\\
 K01925&       68.95800&   1.12&   0.95&   0.88&   2.99&      15.4&    17&  0.658\\
 K02169&	5.45300&   0.97&   0.93&   0.82&   2.24&      43.2&    17&  0.723\\
 K02687&	1.71683&   1.90&   1.94&   1.12&   2.11&      24.2&    17&  0.818\\
 K02720&	6.57148&   0.80&   1.05&   1.05&   3.07&      22.6&    17&  0.619\\

\enddata

\tablecomments{$\ast$: effective combined differential photometric precision~\citep{Jenkins2010b}. $\ast\ast$: correction factor for the bias against planet detection in binary stars. The factor ranges from 0 to 1, with 1 indicating 100\% detection rate even with the flux contamination from a companion star. See \S \ref{sec:bias} for more details. }

\end{deluxetable*}

\clearpage

\begin{deluxetable*}{lcc}
\tablewidth{0pt}
%\tabletypesize{\tiny}
%\setlength{\tabcolsep}{0.02in}
\tablecaption{Ratio of planet occurrence rate between single stars and multiple-star systems as a function of stellar separation.\label{tab:fsfm}}
\tablehead{
\colhead{{$a$}} &
\colhead{{$f_s/f_m$}} &
\colhead{{$\delta f_s/f_m^\ast$}} \\
\colhead{{(AU)}} &
\colhead{{}} &
\colhead{{}} \\
}

\startdata

    1.0&   12.94&   \nodata\\
    2.0&   12.87&   \nodata\\
    5.0&   11.21&   \nodata\\
  10.0&  4.52&  3.16\\
  21.5&  2.99&  1.53\\
  46.4&  2.84&  1.29\\
 100.0&  2.55&  1.01\\
 215.4&  2.43&  0.89\\
  464.2&  2.75&  0.99\\
1000.0&  1.84&  0.57\\
2154.4&  0.87&  0.23\\
4641.6&  0.69&  0.18\\
10000.0&  0.68&  0.18\\

\enddata

\tablecomments{$\ast$: Error bars are calculated based on Poisson statistics and propagated through Equation \ref{eq:fsfm}. No error bar is given within 10 AU because of less than 1 detection of stelar companion according to Equation \ref{eq:nmns}.}

\end{deluxetable*}
\end{document}